\documentclass{jfm}
\usepackage{graphicx}
\usepackage{epstopdf, epsfig}

\usepackage{amsmath}
\usepackage{amstext}
\usepackage{amssymb} 
\graphicspath{{./figures/}}
\usepackage{textcomp}
\usepackage{float}
\usepackage{enumitem}
\usepackage{mathtools}

\usepackage{soul} 

\usepackage{xcolor} 

\shorttitle{Floquet stability of oscillating
boundary layers on adiabatic slopes}
\shortauthor{Kaiser and Pratt}

\title{Floquet stability analyses of oscillating 
boundary layers on adiabatic slopes}

\author{Bryan E. Kaiser\aff{1}
  \corresp{\email{bkaiser@lanl.gov}},
  Lawrence J. Pratt\aff{2}}

\affiliation{\aff{1}Los Alamos National Laboratory, Los Alamos, NM,
USA
\aff{2}Woods Hole Oceanographic Institution, Woods Hole, MA, USA}

\begin{document}

\maketitle

\begin{abstract} 
The presence of a no-slip, impermeable, adiabatic, sloped boundary 
in an otherwise quiescent, stably stratified, Boussinesq flow generates baroclinic vorticity within a diffusive boundary layer. Such conditions are typical of the oscillating boundary layers on adiabatic abyssal slopes, sloped lake bathymetry, and sloped coastal bathymetry in the absence of 
high-wavenumber internal waves, 
mean flows, far-field turbulence on larger scales, and resonant tidal-bathymetric interaction.
We investigate the linear stability of the oscillating 
flow within non-dimensional parameter space typical of the $M_2$ tide and hydraulically smooth, mid-latitude abyssal slopes through Floquet linear stability analysis. 
The flow dynamics depend on three non-dimensional variables: the Reynolds number for Stokes' second problem (Re), the Prandtl number, and a frequency ratio that accounts for the resonance conditions (C, criticality) of the buoyant restoring force and the tidal forcing.
The Floquet analysis results suggest that oscillating laminar boundary layers on 
adiabatic abyssal slopes are increasingly unstable as Reynolds number, criticality 
parameter, and/or spanwise disturbance wavenumber are increased.
We also show that the two-dimensional Floquet linear instability necessarily generates three-dimensional baroclinic vorticity, which suggests that the 
evolution of the gravitational instabilities may be nonlinear as $t\rightarrow\infty$.
\end{abstract}


\section{Introduction}
\begin{figure}
 \centering
 \includegraphics[width=4in]{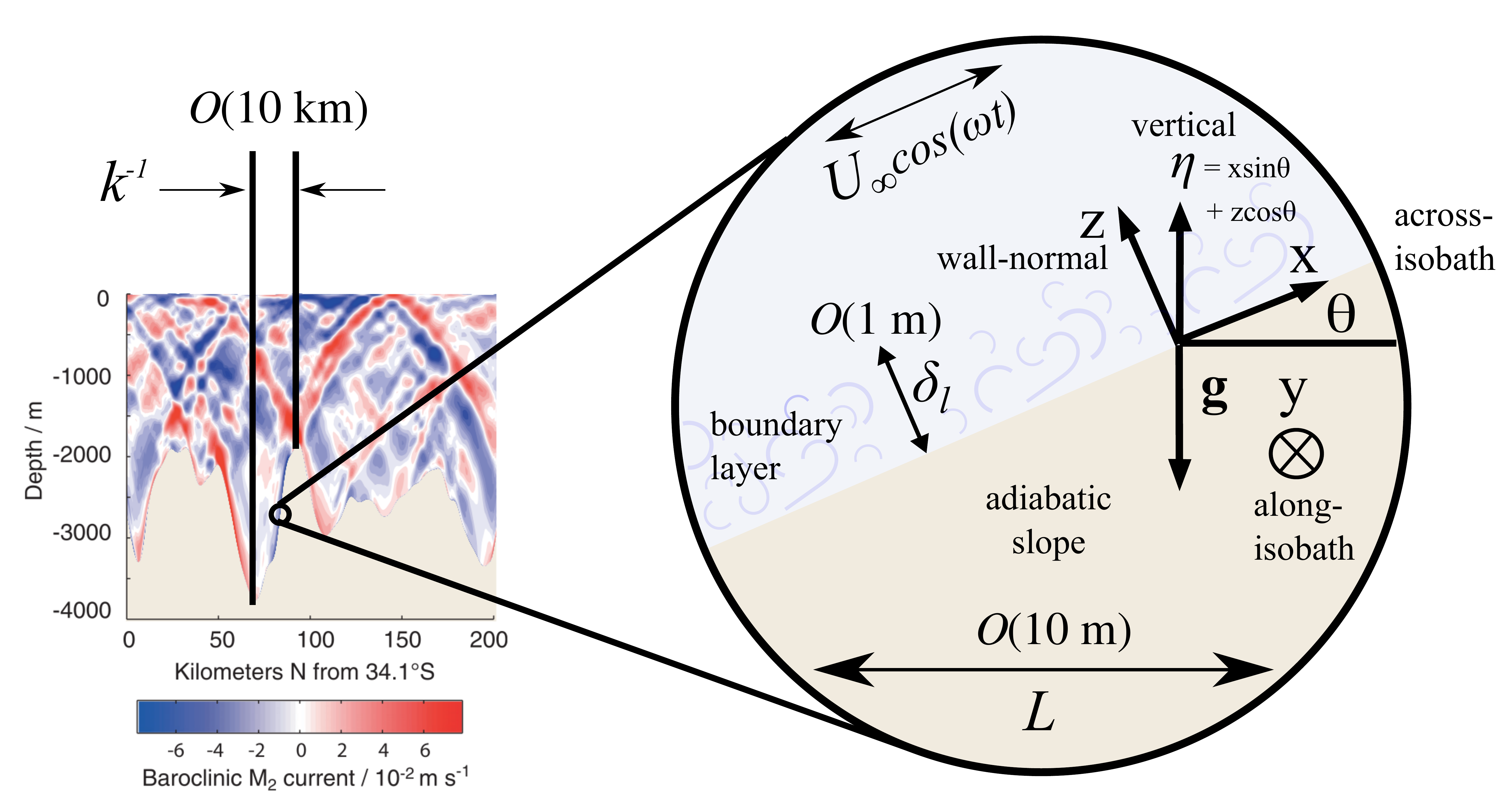}
 \caption{Illustration of Boussinesq boundary layers on abyssal slopes arising from 
 the heaving of density surfaces up and down slope by internal waves 
 with wavelengths $\mathcal{O}(10)$ km. The M$_2$ tide velocity contour plot on the left is 
 from \citet{Zilberman09}.}
 \label{fig:domain}
 \end{figure}

The dynamics of oscillating stratified boundary layers on sloping bathymetry 
may be an important mechanism of diapycnal water mass transformation 
in the context of the global overturning circulation of the ocean
(\citet{Ferrari16}). A fundamental 
understanding of the dynamical pathways between laminar, transitional, and turbulent states is lacking. Historically, analyses of turbulent flows begin
by answering the questions: how stable is the flow to linear disturbances, what are the relevant mechanisms of linear instability, and how does disturbance 
growth change as a function of the relevant 
non-dimensional parameters (\citet{Trefethen93})? We examine the linear 
stability of the laminar boundary layers that form as internal waves 
heave isopycnals up and down infinite slopes to answer these questions for oscillating stratified boundary layers on sloping bathymetry. Our use of a semi-infinite, constant slope 
model is justified by the large separation of 
length scales between the viscous lengthscale of the laminar boundary 
layers, $\mathcal{O}(1)$ cm, and the internal-wave-generative abyssal slope length scales of 
$\mathcal{O}(10)$ km (\citet{Jayne01}, \citet{Goff10}). The geometry and 
scales that are typical
of these boundary layers are illustrated by Figure \ref{fig:domain}.

In the absence of shear, gravitational instabilities are often linear instabilities as is the case for Rayleigh-Taylor and Rayleigh-Benard instabilities. In sheared, gravitationally unstable flows, the transition 
pathways can be more complicated.
Theoretical and experimental evidence 
suggest that 
two dimensional rolls, initiated by linear near-wall gravitational instabilities
and subsequently sheared into 
bursts of three-dimensional turbulence, are the dominant transition-to-turbulence mechanism in gravitationally unstable
Couette flow (\citet{Benard38}, \citet{Chandra38}, \citet{Brunt51}, 
\citet{Deardorff65}, \citet{Gallagher65}, \citet{Ingersoll66}). Recently similar 
mechanisms have been observed in DNS data of oscillating Boussinesq boundary layers 
on adiabatic sloping boundaries,
which are a potentially important regime in the context of abyssal water mass transformation 
(\citet{kaiser2020finescale}).
In this article, the linear stability of 
the non-rotating boundary layers is calculated 
to determine if the gravitational instability 
is a significant linear instability across a much broader 
region of parameter space than can be sampled by direct numerical simulations.

The linear stability of stationary flows is conventionally analyzed
by introducing infinitesimal disturbances to the flow and 
linearizing the governing equations about the stationary base flow to 
form governing equations for the growth or decay of the infinitesimal disturbances (\citet{Trefethen93}).
However, the disturbance equations for oscillating base flows contain
time-periodic coefficients and, therefore, conventional eigenvalue
methods for analyzing the linear growth or decay of disturbances cannot be applied to 
oscillatory flows. 
Instead, the linear stability is determined by applying 
instantaneous instability theory (IIT) or Floquet global instability 
theory (\citet{Luo10}).

IIT is the \textit{ad hoc} application of 
conventional linear instability theory to 
examine the stability of 
the base flow at a discrete time (\citet{vonKerczek76}). 
For example, in the case of Stokes' second problem (SSP, sometimes 
referred to as Stokes layers in the literature),
the Orr-Sommerfeld equation is solved for 
the growth rates of disturbances to 
base flow at a chosen instant
in the period. To evaluate the global stability, or stability 
over the entire period, the instantaneous stability 
calculation must be performed over many instants within 
the period. If one or more instantaneous modes exhibit  
positive growth rates
throughout the period, then the flow is 
globally unstable according to IIT \citet{Luo10}.
However, the validity of the IIT approach rests on 
the assumption that the instantaneous growth rates 
are much larger than the frequency of the base flow, i.e. the 
quasi-steady flow assumption.  IIT is justifiable for the stability 
calculations for high Reynolds number, constant density 
flows because the instantantenous growth rates increase as 
a function of
Reynolds number (\citet{Dwoyer87}).
However, low- to moderate-Reynolds number Stokes layer 
calculations by 
\citet{Luo10} showed that global stability estimates from IIT,
which by definition 
fail to represent linear energy exchanges between instantaneous 
modes, are not predictive of linear global instabilities.
 
Floquet instability theory (\citet{Floquet83}) 
pertains to the net growth or 
suppression of instabilities over the course of 
one period. All periodic instantaneous globally 
unstable 
modes are unstable Floquet modes 
(\citet{Luo10}, \citet{Dwoyer87}), 
but the opposite is 
not true: an unstable Floquet mode can correspond 
to linear energy exchange between two or more 
instantaneous 
modes that do not produce IIT global stability.
Therefore, the evolution of a Floquet mode over a period 
does not necessarily correspond to the 
evolution of an
instantaneous instability that occurs during that period, 
but it does represent the global effect of 
linear instantaneous instabilities.

In this Article, we examine gravitational instabilities 
in laminar oscillating flow on adiabatic slopes in 
which the oscillatory forcing is oriented in the across-isobath 
direction for
parameter regimes typical of super-inertial dynamics in abyssal ocean at low- to 
mid-latitudes.
First, we discuss Floquet analysis of tensors, and, second,
we define the Floquet stability problem. Third, we discuss a 
simple numerical method, and then we discuss the neutral 
stability curves over a broad range of subcritical and supercritical 
slopes before concluding.

\section{Problem formulation}
\begin{figure}
 \centering
 \includegraphics[width=2.0in,angle=0]{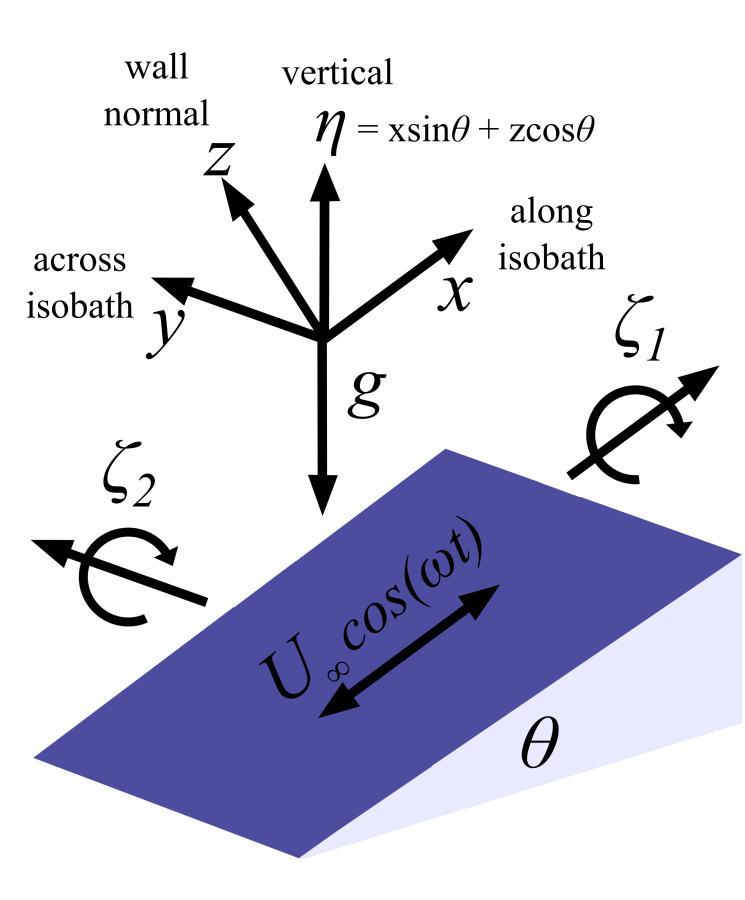}
 \vspace{-5mm}
 \caption{Coordinate system and vorticity components.}
 \label{fig:components}
\end{figure}

Even in the absence of an oscillatory body force, motion arises 
in Boussinesq diffusive boundary layers on adiabatic sloping boundaries 
because baroclinic vorticity is created 
by the tilting density surfaces parallel to the wall normal axis, such that 
the angle $\theta$ separates density surfaces from 
the hydrostatic pressure gradient in the vertical 
within the diffusive boundary layer (see Figure \ref{fig:components}). 
The baroclinic vorticity is oriented 
in the along-slope, constant isobath 
direction ($y$ axis in Figure \ref{fig:components}), 
which drives across-slope wall parallel flows with a net
upslope transport.
\citet{Phillips70} and \citet{Wunsch70} 
simultaneously derived analytical solutions for these laminar 
flows that were validated by the laboratory experiments of \citet{Peacock04}.

The addition of an oscillating
body force in the across-slope direction 
($x$ axis in Figure \ref{fig:components})
gives rise to a class of boundary layers that, in various limits, collapse 
to familiar classical oscillating boundary layers (e.g. Stokes' second problem if the stratification 
vanishes,  
Stokes-Ekman layers in a rotating reference frame, Stokes-buoyancy layers if $\theta=\pi/2$, etc) 
and it is representative of
the frictional interaction of 
low-mode extra-critical baroclinic tidal flows in the ocean. 
\citet{Baidulov10} derived the linear 
solutions for the oscillating, stratified, viscous, and diffusive boundary layer 
in a stationary (not rotating) reference frame 
(hereafter the oscillating boundary layer, OBL) 
and found that the linear flow is a superposition of 
two evanescent modes. \citet{Baidulov10} noted 
that the phase of one of 
the boundary layer modes changes sign
as the slope increases 
from subcritical to supercritical, where 
critical slope is defined by the slope angle $\theta_c$ that 
satisfies $\omega=N\sin\theta_c$ and $N$ is the buoyancy frequency. The 
criticality parameter, defined by dimensional analysis of the governing 
equations, is
\begin{equation}
    \textit{C}=\frac{N\sin\theta}{\omega},\label{eq:C_def}
\end{equation}
where if $\theta=\theta_c$ then $\textit{C}=1$ and subcritical and supercritical 
slopes can be defined as $\textit{C}<1$ and $\textit{C}>1$, respectively.
The change in sign of the boundary layer solution mode 
indicates that the boundary layers 
share some of the dynamics of the parent flow (i.e. the larger scale internal wave 
field in the oceanic example), which undergoes a change of sign of the group velocity of the 
radiated or reflected internal waves 
as the slope angle increases from subcritical to supercritical topography.
At critical slope, the OBL and the far field 
flow resonate because the frequency of buoyant restoring force parallel 
matches that of the across-isobath velocity oscillation.

A commonly observed OBL flow 
feature is the formation and growth of gravitational instabilties 
produced by the upslope advection of relatively heavy water 
over relatively light water trapped at the boundary by friction.
The energy source for OBL gravitational instabilities is the 
baroclinic tide, the same as for 
near boundary
gravitational instabilities and overturning formed by 
critically reflecting internal waves (\citet{Dauxois99}) and 
the nonlinear baroclinic 
tide generation at critical slope (\citet{Rapaka13}, \citet{Gayen11}, \citet{Sarkar17}).
However, gravitational instabilities at critical slope 
are formed by primarily inviscid nonlinearities in the baroclinic 
response to the barotropic tide (\citet{Dauxois99}); 
whereas, the OBL graviational instabilities 
are formed by viscous, insulating boundary conditions 
(\citet{Hart71}). 
OBL gravitational instabilities on extra-critical slopes
have been observed in experiments (\citet{Hart71}),
and observed in OBLs in lakes associated with
internal seiche waves (\citet{Lorke05})  
and internal gravity waves (\citet{Lorke08}).
Similar boundary layer gravitational instabilities have been 
observed 
in the flood (i.e. upslope) phase of estuarine tidal flows 
(\citet{Simpson90}, \citet{Chant01}, \citet{Geyer14}), 
formed
by a combination of bottom friction and the straining
of horizontal buoyancy 
gradients over shallow finite topography.

Supercritical slope OBL laboratory experiments
by \citet{Hart71} identified spanwise
plumes and rolls 
(described by the streamwise, or across-slope vorticity component), 
associated 
with the periodic reversals of the density gradient, 
that 
qualitatively resembled the rolls that appeared
in high Rayleigh number Couette flow experiments by 
\citet{Benard38}, \citet{Chandra38}, and \citet{Brunt51}.
Perhaps due to the similarity to the convection experiments,
the rolls observed by \citet{Hart71} are often referred to
as ``convective rolls'' although the term is misleading 
because it implies diabatic processes are at work;
the gravitational instabilities, rolls, and overturns of interest 
in this study are locally adiabatic. 
Linear stability analyses by \citet{Deardorff65}, \citet{Gallagher65}, 
and \citet{Ingersoll66},
revealed that the observed growth of  
gravitationally unstable disturbances in
high Rayleigh number Couette flows
is suppressed in the plane of the shear (the streamwise-vertical plane) 
by the shear itself
(i.e. the suppression of the spanwise vorticity 
disturbances). 
However, they also found that the growth of disturbances in the spanwise-vertical 
plane (steamwise vorticity 
disturbances) is unimpeded by the shear and grows in the same manner 
as pure convection. 
It has since been established that 
streamwise (the across-isobath direction)
vortices with axes in the 
direction of a mean shear flow (a.k.a. ``rolls'')
can arise due to heating or centrifugal effects (\citet{Hu97}).
Therefore,
since the upslope phase of the OBL is dynamically similar to
gravitationally unstable Couette flow, we hypothesize that 
that linear steamwise vorticity 
disturbances may be an important mode of instability in OBLs.

In this study, we analyze the Floquet stability of the coupled streamwise vorticity component ($\zeta_1$, pointing in the 
across-isobath
$x$ direction in Figure \ref{fig:components}) 
and  buoyancy anomaly (aligned with $-g$ in Figure \ref{fig:components})
in a diffusive, Boussinesq flow on an adiabatic slope when forced by 
an oscillating body force that represents the pressure 
gradient of a low-wavenumber internal wave,
\begin{equation}
 F(t)
 =-A \sin t,
 \label{eq:forcing}
\end{equation}
where 
$A$
is the amplitude of the non-dimensional pressure gradient,
and $x$ is the across-slope coordinate (up/down the slope).
The coordinate system, rotated angle $\theta$ counterclockwise 
from horizontal, and the vorticity components are shown in Figure \ref{fig:components}.

\subsection{Stationary and oscillating base flow components}
The Boussinesq, dimensional form of the  
the conservation equations for mass and momentum are
\begin{equation}
    \nabla\cdot\tilde{\mathbf{u}}=0,
    \label{eq:mass}
\end{equation}
\begin{equation}
    \partial_t \tilde{\mathbf{u}} +\tilde{\mathbf{u}}\cdot\nabla \tilde{\mathbf{u}}
    =-\nabla \tilde{p}+ \nu\nabla^2\tilde{\mathbf{u}}
    +(\tilde{b}\sin\theta+F(t))\mathbf{i}+
    \tilde{b}\mathrm{cos}\theta \mathbf{k},
    \label{eq:momentum}
\end{equation}
\begin{equation}
      \partial_t\tilde{{b}}+\tilde{\mathbf{u}}\cdot\nabla\tilde{b}
  =
    \kappa
    \nabla^2\tilde{b},\label{eq:energy}
\end{equation}
where $\theta$ is the counterclockwise angle of the slope 
from horizontal and the coordinates $\mathbf{i}$, $\mathbf{j}$, 
and $\mathbf{k}$ point in the across-isobath, along-isobath, and 
wall-normal directions. The buoyancy is defined by density anomalies from 
the background flow, $\tilde{b}=g(\rho_0-\tilde{\rho})/\rho_0$, where $g$ is 
the gravitational acceleration, $\tilde{\rho}$ is the anomalous density, and 
$\rho_0$ is the reference density. The pressure $\tilde{p}$ is 
defined as the mechanical pressure divided by the reference density. The buoyancy frequency 
is defined by the hydrostatic background $N=\sqrt{-g/\rho_0(\partial\overline{\rho}/\partial x\sin\theta+\partial\overline{\rho}/\partial z\cos\theta)}$ where 
$\overline{\rho}(x,z)+\rho_0$ is the hydrostatic background density field. 
The domain is semi-infinite, bounded 
only by a sloped wall at $z=0$ with no-slip, impermeable, and 
adiabatic boundary conditions,
\begin{flalign}
    \tilde{\mathbf{u}}(x,y,0,t)&=0,\label{eq:u_bc_wall}\\
    \partial_z\tilde{{b}}(x,y,0,t)&=0.\label{eq:b_bc_wall}
\end{flalign}
At $z\rightarrow\infty$ the flow has two components: a stationary, quiescent, stably stratified, and hydrostatic component and an across-isobath, adiabatic, balanced,
oscillation in the buoyancy, velocity, and pressure fields. These 
two components prescribe the boundary conditions at $z\rightarrow\infty$
\begin{flalign}
    \tilde{\mathbf{u}}(x,y,\infty,t)&={U}_{\scalebox{.9}{$\scriptscriptstyle \infty$}}(t),\label{eq:u_bc_inf}\\
    \partial_z\tilde{{b}}(x,y,\infty,t)&=N^2\cos\theta.\label{eq:b_bc_inf}
\end{flalign}
Boundary conditions for the pressure field are not required because 
the pressure field is diagnosed from the other variables for both flow components.
Let the prognostic variables 
be decomposed linearly into two components, denoted by 
the subscript ``S'' for the stationary component and the primed 
variables for the oscillating component,
\begin{flalign}
  \tilde{\mathbf{u}}(y,z,t)&=
  \mathbf{u}_{\text{S}}(z)+\mathbf{u}'(y,z,t),
  \label{eq:u_decomp}
  \\
  \tilde{b}(x,y,z,t)&=b_{\text{S}}(x,z)+b'(y,z,t),
  \label{eq:b_decomp}
\end{flalign}
The stationary flow has no variability, other than hydrostatically balanced gradients, in the wall-tangent directions such that $\partial_x=\partial_y=0$ 
and contains non-zero velocities only in the across-isobath component 
within a diffusion-driven boundary layer at the wall. The stationary 
flow is itself a linear superposition of a diffusion-driven boundary 
layer component and a quiescent, hydrostatic component, and the 
stationary flow solutions were derived by \citet{Phillips70} and \citet{Wunsch70}. 

Let us decompose the oscillating component into two components:
a base flow component that is the
hydrostatic
response to the across-isobath momentum 
forcing (Equation \ref{eq:forcing}), which includes a 
boundary layer in which friction and the diffusion of the adiabatic 
boundary condition break the inviscid balance of across-isobath heaving of isopycnals, and infinitesimal disturbances to the buoyancy and across-isobath vorticity,
\begin{flalign}
  {\mathbf{u}}'(y,z,t)&=
  \mathbf{U}(z,t)+\epsilon\hat{\mathbf{u}}(y,z,t),
  \label{eq:u_oscil_decomp}
  \\
  {b}'(y,z,t)&=B(z,t)+\epsilon\hat{b}(y,z,t),
  \label{eq:b_oscil_decomp}
\end{flalign}
where $0<\epsilon\ll1$ and
the capitalized variables represent the base flow and 
the hatted variables represent the infinitesimal disturbances. 
The base flow solutions
include only an across-isobath velocity component, and
zero variability in the wall-tangent directions ($x$ and $y$) is also assumed for the base flow
component.

The relationship between the amplitude $A$ of the 
balanced oscillations and the criticality parameter 
reveal that the base flow is an internal wave packet that 
is generated at or reflected by a topographic feature 
when the horizontal length scales of the feature and 
the internal wave are much greater than the relevant 
boundary layer length scales. The governing 
equations for the balanced, inviscid oscillations that 
are set in motion by the momentum forcing can be derived from
Equations \ref{eq:mass} through \ref{eq:energy} by 
assuming no variability in any direction for the momentum and buoyancy 
and by assuming that the flow is adiabatic and quiescent in the 
along-isobath and wall-normal directions everywhere ($V{\scalebox{.9}{$\scriptscriptstyle \infty$}},W{\scalebox{.9}{$\scriptscriptstyle \infty$}}=0$),
\begin{flalign}
  \partial_t{{U}_{\scalebox{.9}{$\scriptscriptstyle \infty$}}}
  &=
  {B}_{\scalebox{.9}{$\scriptscriptstyle \infty$}}\sin\theta
  -A\sin t
  \label{eq:linear_xmomentum_dim}\\
  \partial_t{{B}_{\scalebox{.9}{$\scriptscriptstyle \infty$}}}
  &=-{U}_{\scalebox{.9}{$\scriptscriptstyle \infty$}}N^2\sin\theta, 
    \label{eq:inviscid_linear_dim}
\end{flalign}
The solutions to the balanced, inviscid oscillations 
governed by Equations \ref{eq:linear_xmomentum_dim} 
and \ref{eq:inviscid_linear_dim} are
\begin{flalign}
  U{\scalebox{.9}{$\scriptscriptstyle \infty$}}(t)&=
  -U_0\cos t,\\
  B{\scalebox{.9}{$\scriptscriptstyle \infty$}}(t)&=
  B_0\sin t,
\end{flalign}
which requires a specific relationship between 
the forcing acceleration amplitude $A$, the forcing frequency 
$\omega$, the forcing velocity amplitude $U_0$, and the 
criticality parameter must be satisfied,
\begin{equation}
    A=U_0\omega(\textit{C}^2-1).
    \label{eq:A}
\end{equation}
Either the momentum or buoyancy amplitude $U_0$ or $B_0$ 
can be prescribed so long as the other satisfies
\begin{equation}
\frac{B_0}{U_0}=\textrm{C}N.
\label{eq:amplitude_heaving}
\end{equation}
Equation \ref{eq:A} indicates that the approximation of a 
isopycnal heaving by large horizontal wavelength internal wave on a 
similarly large horizontal wavelength topographic feature breaks down ($A\rightarrow0$)
as the slope angle vanishes $\theta\rightarrow0$ or 
slope-parallel buoyancy oscillations resonate with the forcing $\textit{C}\rightarrow1$, a.k.a. critical slope. This is consistent 
with internal wave theory, which indicates that at critical slope 
balanced, inviscid oscillations become highly nonlinear (\citet{Dauxois99}).

\citet{Baidulov10} derived solutions to for base flow,
\begin{equation}
 U({z},t)=U_0\Real\Big[
     \Big(
     (\alpha_1+
     \text{i}\alpha_2)\mathrm{e}^{(1+\text{i}){z}\phi_1/\delta_1}+     
     (\alpha_3+
     \text{i}\alpha_4)
     \mathrm{e}^{(1+\text{i}){z}\phi_2/\delta_2}     
     -1
     \Big)\mathrm{e}^{\text{i}\omega{t}}
     \Big],
     \label{eq:velocity_sup}
  \end{equation}
\begin{equation}
 B({z},t)=B_0\Real\Big[
  \Big((\beta_1+\text{i}\beta_2)
  \mathrm{e}^{(1+\text{i}){z}\phi_1/\delta_1}  
  +(\beta_3+\text{i}\beta_4)
   \mathrm{e}^{(1+\text{i}){z}\phi_2/\delta_2} 
  -1
  \Big)\text{ie}^{\text{i}\omega{t}}\Big],
  \label{eq:bsoln_oscil_sup}
\end{equation}
which satisfies the boundary conditions shown in 
Equations \ref{eq:u_bc_wall}, \ref{eq:b_bc_wall}, \ref{eq:u_bc_inf}, and 
$\partial_zB(z\rightarrow\infty,t)=0$. 
The real solutions for $U$ and $B$ are split into two sets of solutions corresponding to
the sign of
\begin{equation}
 \phi=\frac{\omega(1+\Pran)}{\nu}
 -\sqrt{\frac{\omega^2(1+\Pran)^2}{\nu^2}+4\frac{\Pran(N^2\sin^2\theta-\omega^2)}{{\nu^2}}}.
 \label{eq:subcrit_requirement}
\end{equation}
If the Prandtl number, 
\begin{equation}
  \Pran=\frac{\nu}{\kappa}, 
\label{eq:Pr_def}  
\end{equation}
is unity then the expression simplifies to 
\begin{equation}
    \phi=\frac{2\omega}{\nu}(1-\textit{C}),
\end{equation}
thus whether the criticality parameter 
$\textit{C}$ is greater than or less than unity determines 
the base flow boundary layer dynamics. In this study $\Pran=1$, and 
the solution coefficients for both subcritical and supercritical flows 
are provided in Tables \ref{tab:sub} and \ref{tab:sup}. 
If $\textit{C}=1$ the 
base flow is an oscillation at the natural frequency of the system. 

\begin{table}
  \begin{center}
\def~{\hphantom{0}}
  \begin{tabular}{llc}
       $\phi_1$ & -1 \\
       $\phi_2$ & -1 \\
       $\delta_1,\delta_2$ & $(\frac{\omega}{4\nu}(1+\Pran)\pm(\frac{\omega^2}{16\nu^2}(1+\Pran)^2
 +\Pran(\frac{N^2\sin^2\theta-\omega^2}{4\nu^2}))^{1/2})^{-1/2}$ \\
       $\alpha_1$ & ${(\omega\delta_1^2-2\nu\Pran^{-1})}/({L_\text{b}\omega\delta_1})$\\
       $\alpha_2$ & 0 \\
       $\alpha_3$ & $(2\nu\Pran^{-1}-\omega\delta_2^2)
 /(L_\text{b}\omega\delta_2)$\\
       $\alpha_4$ & 0 \\       
       $\beta_1$ & $\delta_1/L_b$ \\
       $\beta_2$ & 0 \\  
       $\beta_3$ & $-\delta_2/L_b$ \\
       $\beta_4$ & 0 \\   
       $L_b$ & 
 $(\delta_1-\delta_2)(2\nu\Pran^{-1}+\omega\delta_1\delta_2))
 /({\omega{}\delta_1\delta_2})$\\
  \end{tabular}
  \caption{Solution coefficients for subcritical  
  slopes, $\textit{C}<1$, where $\Pran=1$.  $\delta_1$, $\delta_2$, and $L_b$ have units of length, all others are dimensionless.}
  \label{tab:sub}
  \end{center}
\end{table}
\begin{table}
  \begin{center}
\def~{\hphantom{0}}
  \begin{tabular}{llc}
       $\phi_1$ & -1 \\
       $\phi_2$ & i \\
       $\delta_1,\delta_2$ & $((\frac{\omega^2}{16\nu^2}(1+\Pran)^2
 +\Pran\Big(\frac{N^2\sin^2\theta-\omega^2}{4\nu^2}))^{1/2}
 \pm\frac{\omega}{4\nu}(1+\Pran)\Big)^{-1/2}$ \\
       $\alpha_1$ & $( \frac{\delta_1}{\omega{}L_b^4}-\frac{2\kappa{}}{\delta_1\omega^2L_b^4})
     \big(\omega\delta_1^2\delta_2-2\kappa\delta_2)$\\
       $\alpha_2$ & $( \frac{\delta_1}{\omega{}L_b^4}-\frac{2\kappa{}}{\delta_1\omega^2L_b^4})
 (2\kappa\delta_1+\omega\delta_1\delta_2^2)$\\     
       $\alpha_3$ & $(\frac{2\kappa{}}{\delta_2\omega^2L_b^4}+\frac{\delta_2}{\omega{}L_b^4})
     (2\kappa\delta_1+\omega\delta_1\delta_2^2)$\\
       $\alpha_4$ & $(\frac{2\kappa{}}{\delta_2\omega^2L_b^4}+\frac{\delta_2}{\omega{}L_b^4})
 (2\kappa\delta_2-\omega\delta_1^2\delta_2)$ \\
       $\beta_1$ & $\frac{\delta_2}{\omega{}L_b^4}(\omega\delta_1\delta_2^2
 +2\kappa{}\delta_1)$ \\
       $\beta_2$ & $\frac{\delta_2}{\omega{}L_b^4}({2\kappa{}\delta_2-
 \omega\delta_1^2\delta_2})$ \\  
       $\beta_3$ & $\frac{\delta_1}{\omega{}L_b^4}(\omega\delta_1^2\delta_2-2\kappa{}\delta_2)$ \\
       $\beta_4$ & $\frac{\delta_1}{\omega{}L_b^4}(
 \omega\delta_1\delta_2^2+2\kappa{}\delta_1)$\\   
 $L_b$ & $(\frac
 {(\delta_1^2+\delta_2^2)(\omega^2\delta_1^2\delta_2^2+4\Pran^{-2}\nu^2)}
 {\omega^2\delta_1\delta_2})^{1/4}$ \\ 
  \end{tabular}
  \caption{Solution coefficients for supercritical  
  slopes, $\textit{C}>1$, where $\Pran=1$. 
  $\delta_1$, $\delta_2$, and $L_b$ have units of length, all others are dimensionless.}
  \label{tab:sup}
  \end{center}
\end{table}

\subsection{Ratio of stationary and oscillating time scales}
Across-isobath vorticity disturbances can grow from the 
stationary flow component and/or the oscillating 
base flow component. In this study, Floquet analyses are only 
applied to examine the linear growth of disturbances to the oscillating 
base flow alone because the disturbances are much more rapidly 
modulated by the oscillating flow component than the stationary 
flow component; thus, the stationary flow can be neglected.

A ratio of flow time scales illustrates why, for the 
parameter ranges applicable to the abyssal ocean, the stationary 
flow can be neglect from Floquet analyses of the disturbances.
The wall normal diffusive time scale 
is the relevant characteristic time scale of the 
stationary diffusion-driven flow because the diffusion of the adiabatic boundary condition 
into the interior induces the boundary layer baroclinic vorticity and momentum (\citet{Dell15}). 
Following 
\citet{Dell15}, the time scale
of the non-rotating flow is
\begin{flalign}
 \tau_\kappa&\sim\delta_0^2/\kappa,\\
 &\sim\frac{\sqrt{\Pran}}{N\sin\theta}, 
\end{flalign}
where the boundary layer thickness of the non-rotating 
stationary diffusion-driven boundary layer is
\begin{flalign}
  \delta_0&=\Big(
    \Pran\frac{N^2\sin^2\theta}{4\nu^2}\Big)^{-1/4}.
    \label{eq:rotating_time}
\end{flalign}
Therefore, the modulation ratio (\citet{Davis76}) is
\begin{equation}
  \mathcal{T}\sim\frac{\tau_\kappa}{\tau_\omega}=\frac{\omega\delta_0^2}{\kappa}.
  \label{eq:time_scale}
\end{equation}
In the limit of 
$\mathcal{T}\rightarrow\infty$, the time scale separation between 
the stationary diffusion-driven flow component and the oscillating flow
component indicates that 
the slower stationary diffusion-driven flow component does not modulate 
the faster oscillatory flow component.
If $\mathcal{T}\rightarrow0$, 
the oscillating flow component varies so slowly relative to the 
stationary diffusion-driven flow component that the steady flow component 
may alter the instabilities of the oscillating component.

\begin{figure}
 \centering
 \includegraphics[width=2.75in,angle=0]{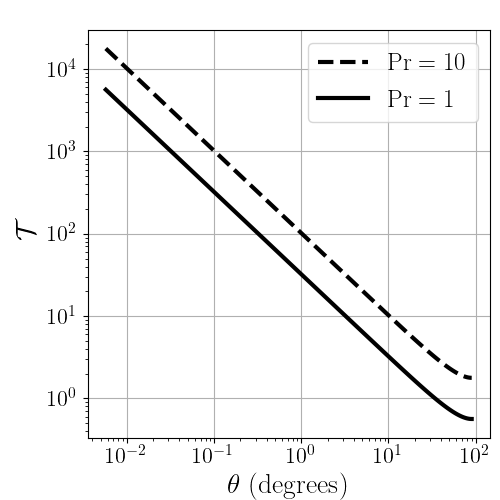}
 \caption{Modulation ratios.}
 \label{fig:time_ratio}
 \end{figure}

The modulation ratios (Equation \ref{eq:time_scale}) for 
typical abyssal parameter ranges for the M$_2$ tide are shown in 
Figure \ref{fig:time_ratio},
which indicates that as $\theta\rightarrow0$, 
the stationary diffusion-driven flow component does not modify 
the much faster dynamics of the oscillatory flow component.
The time scale separation of at least $\mathcal{O}(10^2)$, valid for approximately $0<\theta<1^\circ$ (0.0175 rad), a range of 
slopes commonly found in deep ocean bathymetry (\citet{Goff10}), 
informs our neglect of the stationary diffusion-driven flow component 
and application of Floquet analysis to the oscillatory flow component alone. 
While we consider the case of $\mathrm{Pr}=1$, note that the time scale 
separation increases for $\mathrm{Pr}=10$.

\subsection{Governing equations for spanwise disturbances}
The non-dimensional parameters
of the linearized governing equations for 
the across-isobath vorticity component and buoyancy disturbances
are chosen in the same manner 
as was used \citet{Blennerhassett02} to 
investigate the linear stability of Stokes' second problem, 
with the exceptions that we 
include the 
Boussinesq buoyancy and we analyse the across-isobath (streamwise) 
vorticity instead of the spanwise vorticity.
The oscillating flow variables are non-dimensionalized 
as follows: 
\begin{flalign}
\mathbf{x}=\frac{\mathbf{x}_\text{d}}{\delta}, \hspace{4mm}
{\mathbf{u}}'=\frac{{\mathbf{u}}_\text{d}'}{U_0}, \hspace{4mm}
t={\omega}{t_\text{d}}, \hspace{4mm} 
{p}'=\frac{{p}_\text{d}'}{U_0^2}, \hspace{4mm}
{b}'=\frac{\omega{b}_\text{d}'}{N^2U_0\sin\theta}, 
\label{eq:non_dim_def}
\end{flalign}
where subscript d denotes dimensional variables and 
from here forward the variables without this subscript are 
dimensionless.
${p}'$ is the mechanical pressure divided by the reference density $\rho_0$, and 
the buoyancy is $b'=g(\rho_0-\rho')/\rho_0)$. 
Note that the Eulerian time scale is not 
proportional to the advective time scale 
(i.e. $U_0/\delta\neq\omega$).
The two-dimensional flow governing equations for mass, momentum, and thermodynamic energy (buoyancy) 
for the disturbances 
in the $y-z$ plane are
\begin{flalign}
 0&=\partial_y\hat{v}+\partial_z\hat{w},\\
 \partial_t\hat{v}&=
 -\frac{\Rey}{{2}}\partial_y\hat{p}
 +\frac{1}{2}
 (\partial_{yy}+\partial_{zz})\hat{v},\label{eq:y_momentum}\\
 \partial_t\hat{w}&=
 -\frac{\Rey}{{2}}\partial_z\hat{p}
 +\frac{1}{2}
  (\partial_{yy}+\partial_{zz})\hat{w}
  +\textit{C}^2\hat{b}\cot\theta,\label{eq:z_momentum}\\
 \partial_t\hat{b}&=
 -\Big(\frac{\Rey\partial_zB(z,t)}{{2}}\Big)\hat{w}
 +\frac{1}{2\Pran}
 (\partial_{yy}+\partial_{zz})\hat{b},\label{eq:b_disturbance}
\end{flalign}
where 
the Reynolds number and 
Stokes' layer thickness are
\begin{flalign}
\Rey&=\frac{U_0 \delta}{\nu},\label{eq:Re_def}\\
\delta&=\sqrt{\frac{2\nu}{\omega}}.\label{eq:delta_def}
\end{flalign}
For all analyses in this study $\Pran=1$.
The streamwise vorticity component (see Figure \ref{fig:components}) is defined:
\begin{equation}
 \hat{\zeta}_1=
 \partial_y\hat{w}
 -\partial_z\hat{v}.
\end{equation}
Let the infinitesimal 
disturbances take the form of a normal mode decomposition in the spanwise ($y$) direction:
\begin{flalign}
 \hat{\zeta_1}(y,z,t)&=\zeta_1(z,t)\mathrm{e}^{\text{i}ly} \hspace{1mm} + \hspace{1.5mm} \text{complex conjugate},\\
 \hat{b}(y,z,t)&={b}(z,t)\mathrm{e}^{\text{i}ly} \hspace{1mm} + \hspace{1.5mm} \text{complex conjugate},\label{eq:b_modes}
\end{flalign}
where the modal streamfunction ${\psi}$ and modal velocities are defined
\begin{flalign}
\zeta_1&=(\partial_{zz}-l^2){\psi},
\label{eq:inversion1}\\
 (v,w)&=(-\partial_z\psi,\text{i}l\psi),\\
 l&=l_\mathrm{d}\delta{}=\frac{2\pi}{\lambda}\delta
\end{flalign}
Finally the governing equation for the evolution of the streamwise vorticity modes is
\begin{equation}
 \partial_t\zeta_1=
 \underbrace{\frac{(
 \partial_{zz}-l^2)}{2}{\zeta}_1}_{\text{diffusion}} 
 \hspace{2mm}+\hspace{-3mm}\underbrace{\text{i}l\textit{C}^2b\cot\theta
 }_{\substack{{\text{baroclinic}}\\ {\text{production of vorticity}}}}\hspace{-3mm},
 \label{eq:streamwise_zeta1}
 \end{equation}
and the governing equation for the evolution of the associated buoyancy is
\begin{equation}
 \partial_t{b}=\hspace{1mm}
 \underbrace{\frac{\big(
 \partial_{zz}-l^2\big)}{2\Pran}b}_{\text{diffusion}}\hspace{1mm}
 -\hspace{1mm}\underbrace{\frac{\partial_zB(z,t)\text{i}l\Rey}{{2}}{\psi}}_{\substack{{\text{advection of}}\\ {\text{base buoyancy}}}}.
 \label{eq:buoyancy_zeta1}
\end{equation}
There are no terms representing the advection 
of disturbances by the base flow in
Equations \ref{eq:streamwise_zeta1} and \ref{eq:buoyancy_zeta1}, 
nor the advection of base vorticity by vorticity disturbances. The basic 
state flow enters the equations only through the advection of base buoyancy 
by the buoyancy disturbances (the first term on the right hand side 
of Equation \ref{eq:b_disturbance}); therefore, disturbance vorticity can only be 
produced by gravitational instabilities. 
The state vector for 
Equations \ref{eq:streamwise_zeta1} and \ref{eq:buoyancy_zeta1}
is
\begin{equation}
 \mathbf{x}(z,t)=
 \begin{bmatrix*}
  {\zeta}_1(z,t) \\
  {b}(z,t)
 \end{bmatrix*},
\end{equation}
and the dynamical operator for the evolution of the 
principal fundamental solution matrix (Equation \ref{eq:matrix_equation_pfsm}) is
\begin{equation}
 \mathbf{A}(z,t)=
 \begin{bmatrix*}
  \frac{(\partial_{zz}-l^2)}{2} 
  & \text{i}l\textit{C}^2\cot\theta \\
  -\frac{\partial_zB(z,t)\text{i}l\Rey(\partial_{zz}-l^2)^{-1}}{2} & \frac{(\partial_{zz}-l^2)}{2\Pran} 
 \end{bmatrix*}.
 \label{eq:Astreamwise}
\end{equation}

\subsection{Discrete Floquet exponents, modes, and multipliers}\label{state_vectors}
Let the disturbance 
state vector $\mathbf{x}$ be composed of two variables that vary in a single dimension ($z$), vorticity $\zeta$ and Boussinesq buoyancy $b$, that 
are discretized onto a grid of $N_z$ discrete points in the $z$ coordinate takes 
the form
\begin{equation}
 \mathbf{x}(t)=
 \begin{bmatrix*}
  \zeta(z_1,t) \\
  \vdots \\
  \zeta(z_{N_z},t) \\
  b(z_1,t) \\
  \vdots \\
  b(z_{N_z},t)   
 \end{bmatrix*}
 =
  \begin{bmatrix*}
  x_{1}(t) \\
  \vdots \\
  x_M(t)    
 \end{bmatrix*},
\end{equation}
where the length of state vector $\mathbf{x}$ is $M=N_v\times N_z$ 
(the number of variables multiplied by the number of grid points).
The principal fundamental solution matrix $\boldsymbol{\Phi}_p(t)$ 
for state vector $\mathbf{x}(t)$ is defined as
\begin{flalign}
\boldsymbol{\Phi}_p(t)
  &=
\begin{bmatrix*}
  x_{1,1}(t) & \ldots & x_{1,M}(t)\\
  \vdots & \ddots & \vdots \\
  x_{M,1}(t) & \ldots & x_{M,M}(t)
  \end{bmatrix*}.
  \label{eq:final_modes}  
\end{flalign}
The principal fundamental solution matrix at time $t=T$ ($T$ being the oscillation 
period) can 
be analyzed to determine the fastest growing Floquet mode and 
discrete grid location of the largest Floquet multiplier (see 
Appendix \ref{appA} for the derivation and 
formal properties of the principal fundamental solution matrix). 
At time $t=0$ the principal fundamental solution matrix is an identity 
matrix 
that can be physically interpreted as
a set of independent linear perturbations of the system;
thus, each Floquet mode 
represents an initial disturbance at each discrete grid location.

The innovation of \citet{Floquet83} was the recognition that, 
without a loss of generality, the initial conditions 
specified at time $t=t_0$ 
can be expressed in terms of the eigenvectors of the principal fundamental solution 
matrix after one oscillation period has elapsed, $t=t_0+T$. Choosing $t_0=0$,  
\begin{equation}
  \boldsymbol{\Phi}_p(T)\mathbf{v}(0)=\boldsymbol{\mu}\mathbf{I}\mathbf{v}(0)
\end{equation}
where $\boldsymbol{\mu}$ is a vector of the eigenvalues of $\boldsymbol{\Phi}_p(T)$ and 
$\mathbf{v}$ are the eigenvectors of $\boldsymbol{\Phi}_p(T)$ such that
$\mathbf{v}(0)=\mathbf{x}(0)$ and therefore $\mathbf{v}(T)=\mathbf{x}(T)$. 
$\mathbf{v}(t)$ are the Floquet 
modes and $\boldsymbol{\mu}$ are the Floquet multipliers. 
Therefore, 
\begin{flalign}
 \mathbf{v}(T)=\boldsymbol{\mu}\mathbf{I}\mathbf{v}(0),
 \label{eq:forward_propogator2}
\end{flalign}
is equivalent to Equation \ref{eq:forward_propogator1}.
The stability of the system in terms of Floquet multipliers is:
\begin{enumerate}[label=(\alph*)]
\item If \textit{all} Floquet multipliers (i.e. eigenvalues) satisfy $\Real[{\mu}]<1$, 
then
 all disturbances decay 
 as $t\rightarrow\infty$ 
 and the system is 
 \textbf{stable}. 
\item If \textit{any} Floquet multipliers satisfy $\Real[{\mu}]=1$ 
and the rest satisfy $\Real[{\mu}]<1$, then 
 then the stability of the system 
 is \textbf{periodic} as $t\rightarrow\infty$. Periodic modes do not necessarily oscillate at the base freqency, 
 only if $\mu\pm1+0$i then the mode's frequency exactly matches 
 the base flow.
\item If \textit{any} Floquet multiplier satisfies $\Real[{\mu}]>1$, then 
 the disturbance
 will grow in amplitude as $t\rightarrow\infty$
 and the system is \textbf{unstable}.
\end{enumerate}

Floquet multipliers are generally complex.
For fluid flows, however, the component of interest is the real part of 
$\mu$. The Floquet solutions are the columns of the principal fundamental 
solution matrix while the Floquet modes are defined by the 
Floquet exponents
\begin{equation}
 \boldsymbol{\gamma} = \frac{\log\boldsymbol{\mu}}{T},
 \label{eq:expo}
\end{equation}
where exponents are complex. The Floquet modes are defined 
\begin{equation}
 \mathbf{v}(t)=\mathrm{exp}({\Real[\boldsymbol{\gamma}]t})\mathbf{P}(t)\mathbf{v}(0),
\end{equation}
thus 
\begin{equation}
    \Real[\boldsymbol{\Phi}_p(T)]=\mathrm{exp}({\Real[\boldsymbol{\gamma}]T}),
\end{equation}
where $\mathbf{P}(t)$ are periodic Floquet mode components (harmonics of the base frequency)
and the Floquet multipliers can be expressed as $\boldsymbol{\mu}=\mathrm{exp}({\Real[\boldsymbol{\gamma}]T})$ because $\mathbf{P}(T)=\mathrm{exp}({\Imag[\boldsymbol{\gamma}]T})=1$.
The real part of the Floquet exponents corresponds to the 
growth or decay of the mode as $t\rightarrow\infty$ 
and the imaginary part of the Floquet 
exponent determines the frequency of the Floquet mode in terms of harmonics of 
the base freqency. Further Floquet theory details are provided in Appendix A.

\subsection{Boundary conditions}
The oscillatory forcing was imposed by imposing a ``moving wall'' boundary condition
rather than applying a body force directly on the evolving modes. 
At the moving wall, the total flow boundary conditions on the momentum 
are no-slip and impermeable; therefore, at $z=0$
\begin{equation}
 \mathbf{u}'
 =U\mathbf{i}+\epsilon\hat{\mathbf{u}}=\cos{t}\mathbf{i},
\end{equation}
where $0<\epsilon\ll1$ is a small parameter and
\begin{equation}
 U(0,t)=\cos{t},
\end{equation}
therefore
\begin{equation}
 \partial_{z}{\psi}=0,
 \label{eq:noslip}
\end{equation}
is required to satisfy the no-slip condition 
at $z=0$ for either definition of the streamfunction. 
The 
streamfunction must be constant along an 
impermeable wall; therefore it is numerically convenient to choose
\begin{equation}
 \psi=0,
 \label{eq:imperm}
\end{equation}
at $z=0$ to satisfy
$w=\partial_x\psi=0$ for the spanwise vorticity - streamfunction approach
or $w=\partial_y\psi=0$ for the streamwise vorticity - streamfunction approach.

The wall is adiabatic; therefore,
\begin{equation}
  \partial_z{b}'
 =\partial_zB+\epsilon\partial_z\hat{b}=0.
\end{equation}
Since the basic state stratification satisfies
\begin{equation}
 \partial_zB=0,
\end{equation}
then the disturbance stratification must satisfy
\begin{equation}
 \partial_z\hat{b}=0,
 \label{eq:adia0}
\end{equation}
at $z=0$.

At $z\rightarrow\infty$, the conventional boundary conditions 
for Stokes' second problem are parallel and irrotational flow. Parallel 
flow is ensured if 
\begin{equation}
 \psi=0,
 \label{eq:parallel}
\end{equation}
at $z\rightarrow\infty$. Irrotational flow at $z\rightarrow\infty$ 
is prescribed by
\begin{equation}
 \zeta_1=0. 
 \label{eq:irrot}
\end{equation}
The background stratification is not adiabatic in the 
far field. but the disturbance stratification can be adiabatic 
because the basic flow gradients exist only in the boundary layer. 
Therefore, at $z\rightarrow\infty$
\begin{equation}
 \partial_zb=0.
 \label{eq:adiainf}
\end{equation}

\section{Numerical methods}
As described in section \ref{state_vectors}, the 
length of the perturbation state vector $\mathbf{x}$
is $N_v\times{}N_z$, 
where $N_v$ is the 
number of variables 
and $N_z$ is the number of grid points. 
In the streamfunction-vorticity formulation in this study $N_v=2$ for
vorticity and buoyancy. The number of grid points in $z$, the 
wall-normal direction, for all calculations was $N_z=200$.
Therefore the discrete principal fundamental solution matrix 
is a square matrix with $2N_z$ rows and columns.
The variables were computed at the cell centers of 
a uniform grid of height $H/\delta=32$, where the non-dimensional 
grid encompassed $z=[0,H/\delta]$. Previous studies 
found that Floquet stability calculations for Stokes' 
second problem were unaffected by 
an upper domain boundary as long as it was located at $H/\delta=32$
or greater (\citet{Blennerhassett06}, \citet{Luo10}). 

Centered second-order finite difference schemes were 
used to compute the discrete forms of 
all first and second derivatives
and the vorticity inversions that appear the 
dynamical operator $\mathbf{A}(z,t)$ for 
(Equation \ref{eq:Astreamwise}). 
To implement the no-slip, impermeable
boundary conditions at the wall
(Equations \ref{eq:noslip}, \ref{eq:imperm}),
the streamfunction and its $z$ derivative were set to zero. 
However, to guarantee unique solutions at second-order 
accuracy, the vorticity at the wall was required to 
compute the second derivatives of the vorticity. 
The second-order accuracy was confirmed with 
grid convergence tests shown in Appendix B.
A second-order accurate extrapolation of 
the vorticity at the wall that
accounts for no-slip 
and impermeable boundary conditions 
was derived by (\citet{Woods54}) for 
this purpose. The \citet{Woods54} boundary condition is:
\begin{equation}
 \zeta(0,t)=\frac{3}{\Delta{z}^2}\psi(z_1,t)-\frac{1}{2}\zeta(z_1,t),
\end{equation}
where $z=0$ denotes variables located at the wall and 
$z=z_1$ denotes variables located at the first 
cell center. All of the other boundary conditions 
(Equations \ref{eq:adia0}, \ref{eq:parallel}, 
\ref{eq:irrot}, and \ref{eq:adiainf}) were readily implemented 
into the discrete derivatives within the discrete form of the operator $\mathbf{A}(z,t)$. Finally, test functions 
were used to ensure that the truncation error for all discrete 
derivative and inversions decreased 
with $(\Delta{z})^{-2}$, where $\Delta{z}$ is the 
height of a grid cell. 

To obtain the principal fundamental solution matrix 
at time $t=T$, Equation \ref{eq:matrix_equation_pfsm} 
was integrated over one period with the standard explicit
fourth order Runge-Kutta time advancement method, equivalent 
to simultaneously 
solving the evolution of the state vector in Equation \ref{eq:gov} 
in which each linearly independent solution begins with 
each linearly independent initial condition as defined in Equation \ref{eq:pfsm}. 
The method for computing 
the principal fundamental solution matrix in this study 
is formally second-order accurate. 
\begin{figure}
 \centering
 \includegraphics[width=0.8\textwidth,angle=0]{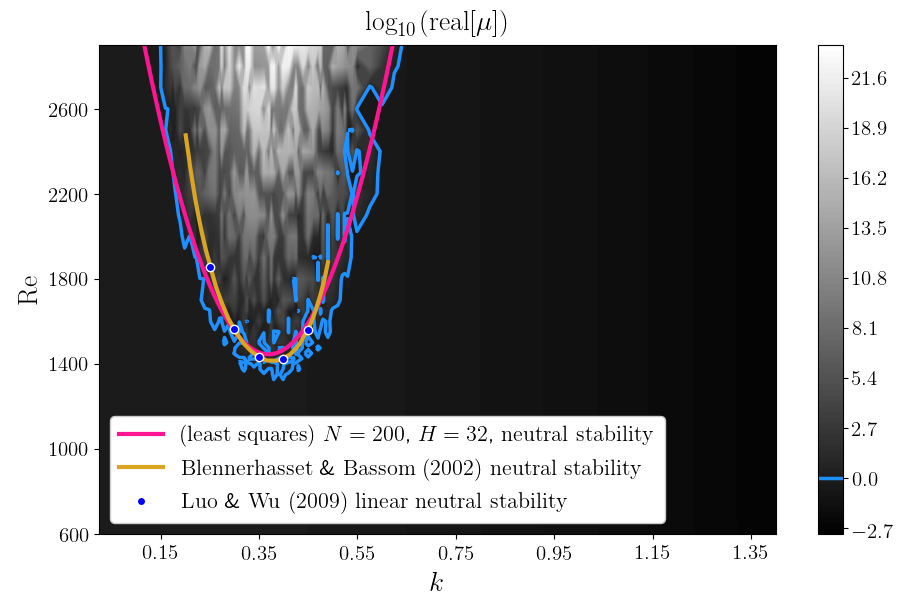}
 \caption{The neutral stability curve for Stokes' second problem.}
 \normalsize{\noindent Verification of spatial discretization, temporal 
 discretization, and eigenvalue calculation.
 The computed Floquet multipliers are shown by 
 the gray shading, and the calculated $\Real[\mu]=1$ contour is 
 represented with the blue line.}
 \label{fig:stokes_curve}
\end{figure}

\subsection{Solver verification}
The neutral stability curve for Stokes' second problem was computed 
as a code verification test, shown in Figure \ref{fig:stokes_curve}. 
There 
the blue line is the computed neutral stability curve, and the 
pink line is a least squares fit of the computed neutral stability curve.
The yellow line is a least squares fit of the computed stability 
curve by \citet{Blennerhassett02}, 
who used a spectral method for the computation, and 
the blue dots were calculated by linearized direct numerical 
simulations by \citet{Luo10}. The variations in the 
neutral stability curve about the pink line can be 
attributed to the spatial discretization method. 
This hypothesis is supported by the Floquet 
results for the neutral stability 
curve for Mathieu's equation (see Appendix \ref{appB})
, which has no 
spatial derivatives and was computed to graphical 
accuracy using the same code for time integration 
and eigenvalue calculation.

\begin{figure}
 \centering
 \includegraphics[width=0.7\textwidth,angle=0]{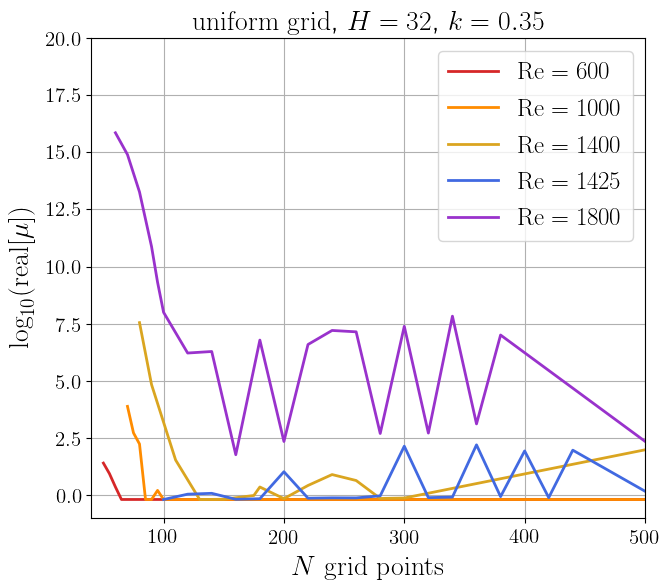}
 \caption{Grid convergence occurs only for stable calculations.}
 \normalsize{\noindent Stable Floquet multiplier 
 calculations achieve grid convergence. Near the 
 neutral stability curve, the Floquet multiplier calculations fail to acheive grid 
 convergence because finite differences of 
 the identity matrix initial condition 
 of the principal fundamental solution matrix introduce grid independent 
 noise. Therefore the ``wiggles'' of the blue curve in Figure \ref{fig:stokes_curve}
 are due to the sensitivity of the multiplier value 
 to noise introduced at just after $t=0$ when 
 finite differences are taken of discontinuities in the principal fundamental 
 solution matrix.
 }
 
 \label{fig:grid_resolution}
\end{figure}

The irregularities of the blue neutral stability curve 
in Figure \ref{fig:stokes_curve} occur because of 
the initialization of the principal fundamental solution 
matrix as an identity matrix. In the first time step 
of the calculation, the finite differencing 
of discontinuous functions (specifically Dirac delta functions)
introduces discretization errors that do not converge 
with increased grid resolution. To check this, 
the stability of Stokes' second problem was computed for varied 
Reynolds number and grid resolution at $k=0.35$, 
as shown in Figure \ref{fig:grid_resolution}. 
The Floquet multipliers in Figure \ref{fig:grid_resolution}
that correspond to stable points in the 
neutral stability plot of Figure \ref{fig:stokes_curve}
converge quickly with increasing grid resolution. 
However, for $\Rey\geq1400$, $k=0.35$ (inside 
the unstable region of Figure \ref{fig:stokes_curve}) the 
multipliers in Figure \ref{fig:grid_resolution} 
do not converge with increasing grid resolution. 
\citet{Luo10} pointed out that SSP stability calculations 
are extremely sensitive to transient noise that occurs 
during the course of the oscillation, which suggests 
that small round-off errors and other numerical noise may 
explain the 
variations of the stability curve fit calculated by 
\citet{Blennerhassett02}. Figure \ref{fig:grid_resolution} 
indicates that the primary culprit for transient noise 
in the present study is the introduction of 
discretization errors at the first time step. Therefore 
the $N_z=200$ was deemed sufficient 
grid resolution, and the calculated neutral stability 
curves from a finite difference method must be considered 
approximate rather than exact.

\section{Results}
\begin{figure}
 \centering
 \includegraphics[width=5.5in,angle=0]{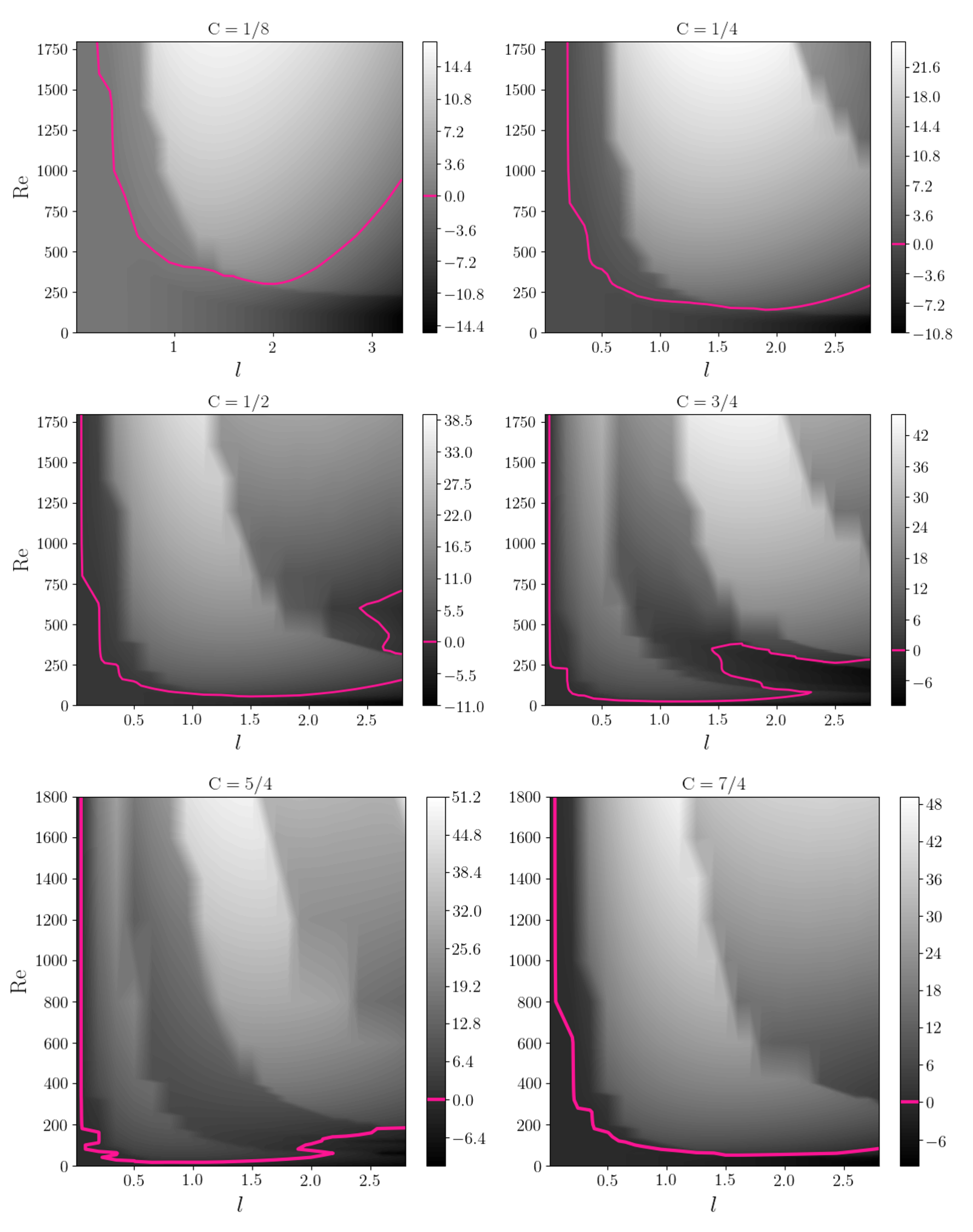}
 \caption{$\log_{10}(\Real[\mu])$ for the streamwise vorticity, $\zeta_1$, for subcritical and supercritical slopes as functions of 
 criticality parameter ($\textit{C}=N\sin\theta/\omega$), Reynolds number ($\Rey=U_0\delta/\nu$), and spanwise disturbance wavenumber 
 ($l=2\pi\delta/\lambda$).
  The pink lines are the Floquet neutral stability curves 
  ($\Real[|\mu|]=1$, $\Real[|\gamma|]=0$).  
 }
 \label{fig:zeta1_subcritical}
\end{figure}
Figure \ref{fig:zeta1_subcritical} depicts the results of Floquet analysis of vorticity and buoyancy disturbances
governed by Equations \ref{eq:streamwise_zeta1} and \ref{eq:buoyancy_zeta1}, 
respectively, which indicate that the system
is increasingly linearly unstable as 
the Reynolds number, criticality parameter, and non-dimensional spanwise 
disturbance 
wavenumber are increased.
$\Rey<10$ and $l<0.5$, approximately, are maximum Reynolds number and 
disturbance wavelength required for stability for the entire investigated parameter space.
At $\textit{C}\ll1$ the 
minimum Reynolds number necessary for global linear instability increases, and
the forcing of vorticity disturbances by buoyancy disturbances decreases because 
baroclinic production term on the right hand side of Equation \ref{eq:streamwise_zeta1} vanishes in the limit $\textit{C}\rightarrow 0$.
Therefore, increased stability at small $\textit{C}\ll1$ suggests 
that baroclinic production of disturbance vorticity is the primary mechanism of instability. 
This result is in agreement with the empirical and approximate stability criteria of \citet{Hart71} 
which posits that 
the flow is globally stable if $\textit{C}^2\ll 1$. 

If $\omega$, $N$, and $\theta$ constant, 
Equation \ref{eq:amplitude_heaving} indicates that
increasing the oscillation velocity amplitude $U_0$ and thus  
the Reynolds number will increase the amplitude of the buoyancy oscillations $B_0$
as well. Since the boundary layer thicknessness $\delta_1,\delta_2$ do not 
depend on $U_0$ and they determine the length scale of the boundary layer buoyancy gradient, the boundary layer buoyancy gradient increases with increasing Reynolds number and both quantities force buoyancy disturbances through the 
second term on the right hand side of Equation \ref{eq:buoyancy_zeta1}. 

The system is stable to large spanwise wavelength disturbances because 
all of the terms on the right 
hand sides of the 
disturbance 
Equations \ref{eq:streamwise_zeta1} and \ref{eq:buoyancy_zeta1} 
vanish except for the diffusion of disturbances in the wall normal direction 
in the limit as $l\rightarrow0$. Growing disturbances described 
by Equation \ref{eq:Astreamwise} are confined 
to the boundary layer because they are forced by the base flow buoyancy gradient that vanishes outside the boundary layer; disturbances that propagate 
outside of the boundary layer are diffused.

The fingers on the low Reynolds number, higher $l$ portions 
of the neutral stability curves shown in Figure \ref{fig:zeta1_subcritical} at $\textit{C}=1/2$ and $\textit{C}=3/4$ 
merit further examination. However, 
irregularities 
in neutral stability curves have also been found in the neutral stability curve for Stokes' second 
problem (\citet{Blennerhassett02}). 

\begin{figure}
 \includegraphics[width=1\textwidth,angle=0]{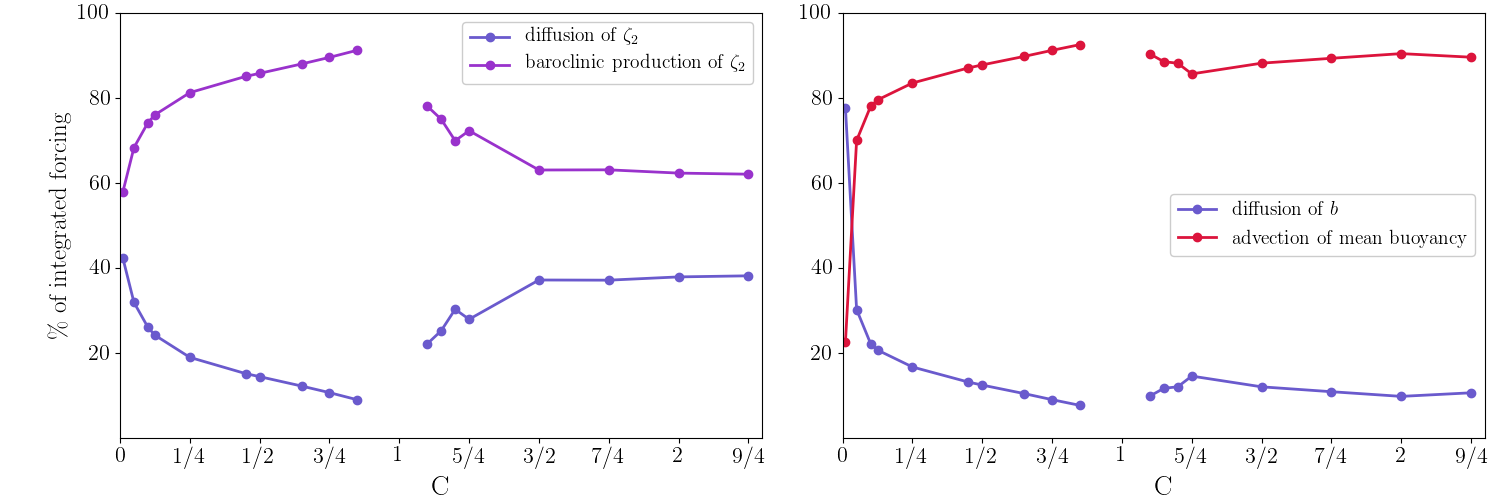}
 \caption{Integrated $\zeta_1$, $b$, budgets for $\Rey=420,l=1.0$.}
  \normalsize{\noindent 
  Percentages of the right hand sides of 
  Equations \ref{eq:streamwise_zeta1} and \ref{eq:buoyancy_zeta1}
  calculated by spatio-temporal 
  integration of $\mathbf{A}(z,t)$, for the most unstable Floquet mode at 
  each value of $\textit{C}$. The flow is Floquet 
  unstable at $\Rey=420,l=1.0$ for all C investigated, therefore 
  the dominant mechanisms induce disturbance growth.
 }
 \label{fig:integrals_streamwise}
\end{figure}

Each equation term in matrix $\mathbf{A}(z,t)$ (Equation \ref{eq:Astreamwise})
was integrated in $z$ and $t$ for the fastest growing Floquet mode 
(the mode with maximum $\mu$, where $\mu>1$)
to qualitatively 
assess the dominant physical mechanisms of the fastest growing linear disturbances.
The terms were normalized by 
the total right hand side forcing, corresponding to the rows of 
$\mathbf{A}(z,t)$, such that the temporally and spatially integrated 
forcings sum to unity. The integrated equation term results are shown in
Figure \ref{fig:integrals_streamwise} suggests that,
on both subcritical and supercritical slopes, 
the fastest growing linearly unstable disturbances 
are amplified by the non-diffusive terms in 
Equations \ref{eq:streamwise_zeta1} and \ref{eq:buoyancy_zeta1}, 
and that the diffusive terms increasingly inhibit the growth of linear instabilities at $\textit{C}\ll1$.

The inherent three dimensionality of the gravitational instability 
initiated by spanwise buoyancy disturbances (Equation \ref{eq:b_modes}) 
to the streamwise vorticity component
is evident
in the total baroclinic vorticity production term that can be derived 
from by linearizing and taking the curl of the momentum equation (Equation \ref{eq:momentum}), 
\begin{equation}
 \text{total baroclinic vorticity production}
 =\textit{C}^2\big(\text{i}lb(z,t)\mathrm{e}^{\text{i}ly}\cot\theta\mathbf{i}
 -\partial_zB\mathbf{j}
 -\text{i}lb(z,t)\mathrm{e}^{\text{i}ly}\mathbf{k}\big).
 \label{eq:baroclinic_production2}
 \end{equation}   
Equation \ref{eq:baroclinic_production2} reveals that spanwise buoyancy
disturbances force the streamwise and wall-normal vorticity components.
This suggests that the growth of linear gravitational instabilities induced by spanwise 
disturbances must induce
three dimensional motion, a phenomena that is widely observed 
in other stratified shear flow instabilities (\citet{Peltier03}), and it 
suggests that the full three-dimensional dynamics must be considered to 
accurately assess the growth of linear instability as $t\rightarrow\infty$.
Comparison of the Floquet analysis results with
the downslope relaminarization of initially two-dimensional spanwise rolls 
that evolve into three-dimensional motion 
observed in experiments (\citet{Hart71}) suggests that the Floquet analysis of 
spanwise disturbances
may be descriptive of the initial linear gravitational instability that triggers 
spanwise rolls, but fails to account for the three-dimensional diffusive damping.
The Floquet analysis identifies linear gravitational instabilities that 
exhibit net growth over the period of the oscillation despite the transience of the unstable buoyancy gradient. However, since Equation \ref{eq:baroclinic_production2} indicates that the instabilities must become 
three dimensional, the growth of these linear instabilities could be suppressed or 
increased by three-dimensional motion. Direct numerical simulations by \citet{kaiser2020finescale} suggest that low Reynolds number the growth of linear instabilities is dissipated by three-dimensional motion.


\section{Conclusions}
Floquet linear stability theory was applied to 
laminar, oscillating, stratified, viscous diffusive boundary layers 
on infinite slopes in non-rotating reference frames. 
The linear stability of a two-dimensional disturbances in the 
$y-z$ plane (described by the streamwise vorticity $\zeta_1$) was
evaluated within the non-dimensional parameter 
ranges of $0<\Rey\leq1750$, $1/8\leq \textit{C}\leq7/4$, $0<l\leq3$, and 
$\Pran=1$. The parameter regime is consistent with 
idealized $M_2$ tidal heaving of isopycnals up and down 
smooth, mid-latitude abyssal slopes where $N/\omega\sim7.1$, 
$0<\theta\leq\pi/12$, $\nu\sim2.0\cdot10^{-6}$ m$^2$s$^{-1}$, 
$\omega\sim1.4\cdot10^{-4}$ rad s$^{-1}$, $35.4\leq\lambda<\infty$ cm, 
and $0.0<U_\infty\leq2.1$ cm s$^{-1}$. 

The most salient results the Floquet analyses are
\begin{itemize}
    \item \hspace{1mm} we have shown that oscillating laminar boundary layers on 
    adiabatic abyssal slopes are increasingly linearly unstable 
    as Reynolds number, criticality parameter, and/or spanwise disturbance 
    wavelength are increased;
    \item \hspace{1mm} the mechanism of the instability is the same as for gravitationally 
    unstable Couette flow;
    \item \hspace{1mm} our results qualitatively agree with experiments of related flows;
    \item \hspace{1mm} the growth of unstable Floquet modes must generate three-dimensional 
    motions that are unaccounted for in the Floquet analyses which could suppress or increase instabilities as $t\rightarrow\infty$;
    \item \hspace{1mm} our finite difference numerical approach is more sensitive to 
    numerical noise in the parameter space near the neutral stability curve 
    than spectral methods.
\end{itemize}


The linear disturbances must create three-dimensional baroclinic vorticity 
production. Therefore the growth of disturbances predicted by two-dimensional Floquet analysis as $t\rightarrow\infty$ may not be descriptive of the the 
three-dimensional flow as $t\rightarrow\infty$. However, the gravitational 
instability described by two-dimensional Floquet analysis is consistent with 
two-dimensional rolls in the $y-z$ plane observed in experiments (\citet{Hart71}), 
which suggests that the Floquet analysis results
describe the initial vorticity and buoyancy mechanisms of
linear gravitational instabilities on Boussinesq adiabatic slopes but not 
the behavior of disturbances as $t\rightarrow\infty$.

Our results validate the hypothesis that 
Floquet stability 
calculations, regardless of the chosen numerical method, are extremely sensitive to 
transient numerical 
noise that occurs throughout the oscillation period, in agreement with the conclusions of \citet{Luo10} for 
Stokes' second problem calculations. The sensitivity to the 
numerical errors and noise may explain the highly irregular  
shapes of neutral stability curves that vary from one 
study to the next (\citet{Blennerhassett02}, \citet{Luo10}).
While the transient noise can represent 
actual physics that perturb disturbances randomly throughout the 
phase of the oscillation,
the statistical characteristics of the numerical noise 
are not readily discernable;
therefore, the disturbance 
conditions of the numerical calculations are
impossible to exactly replicate in laboratory experiments (\citet{Luo10}). 
It was shown that the transient noise cannot be 
eliminated by increasing the numerical accuracy;
therefore, the results of this study support the 
conclusion that numerical 
Floquet stability calculations must be considered 
approximate, rather 
than exact, estimates of linear disturbance behavior.

\section{Acknowledgements}
B.K. was supported by a N.S.F. 
Graduate Research Fellowship and 
the Massachusetts Institute of Technology - Woods Hole 
Oceanographic Institution Joint Program, and
by the National
Science Foundation (OCE-1657870).
The authors 
thank the Massachusetts Green Computing Center, J{\"o}rn Callies, 
Jesse Canfield,
Raffaele Ferrari, 
Karl Helfrich, and
Andreas Thurnherr. This document is approved for Los Alamos Unlimited Release, LA-UR-21-28221.

\appendix

\section{Floquet theory applied to tensors}\label{appA}
In this section
Floquet theory 
for tensors is briefly summarized to show how Floquet multipliers can 
be calculated directly for base flow solutions across spatial grids.
The reader is referred to \citet{Iooss12} for the complete derivation. 

In Floquet theory for vectors or tensors,
the principal fundamental solution matrix is a mapping 
of the state vector $\mathbf{x}$ at time $t=0$ to one 
period, $t=T$. 
Consider the non-autonomous system, 
\begin{equation}
 \frac{\mathrm{d}\mathbf{x}}{\mathrm{dt}}=\mathbf{A}(t)\mathbf{x}(t),
 \label{eq:gov}
\end{equation}
where $\mathbf{x}(t)$ is a vector and the operator $\mathbf{A}(t)$ is periodic
\begin{equation}
 \mathbf{A}(t+T)=\mathbf{A}(t).
\end{equation}
For a state vector $\mathbf{x}(t)$ of shape $[M\times1]$, where
$M$ is the number of variables times the number of grid points, there 
exists a fundamental solution matrix $\boldsymbol{\Phi}(t)$ of 
shape $[M\times{}M]$ and coefficient vector $\mathbf{c}$ of shape $[M\times1]$,
such that
\begin{flalign}
 \mathbf{x}(t)&=\boldsymbol{\Phi}(t)\mathbf{c} \label{eq:gen_soln1}. 
\end{flalign}
The fundamental solution matrix is a non-unique 
matrix 
in which the columns are the structure 
of the linearly independent 
solutions.
The magnitude of the elements in $\boldsymbol{\Phi}$
depend on the choice of $\mathbf{c}$, and the only restriction to 
the choice of a tenable $\mathbf{c}$ is that $\boldsymbol{\Phi}$
be invertible.
In that case, at time $t=0$, 
\begin{equation}
 \mathbf{c}=\boldsymbol{\Phi}(0)^{-1}\mathbf{x}(0)
 \label{eq:obtain_c1}
\end{equation}
Substitution of Equation \ref{eq:obtain_c1} into Equation \ref{eq:gen_soln1} yields
\begin{flalign}
 \mathbf{x}(t)&=\boldsymbol{\Phi}(t)\boldsymbol{\Phi}(0)^{-1}\mathbf{x}(0).
 \label{eq:forward_propogator0}
\end{flalign}
The principal fundamental solution matrix $\boldsymbol{\Phi}_p$ is just 
a fundamental solution matrix chosen such that at $t=0$ it is an identity matrix:
\begin{equation}
 \boldsymbol{\Phi}_p(0)=\mathbf{I}.
 \label{eq:pfsm}
\end{equation}
Substitution of Equation \ref{eq:pfsm} in Equation \ref{eq:forward_propogator0} yields
\begin{flalign}
 \mathbf{x}(T)=\boldsymbol{\Phi}_p(T)\mathbf{x}(0);
 \label{eq:forward_propogator1}
\end{flalign}
therefore, the principal fundamental solution matrix at time $t=T$ maps 
the initial state $\mathbf{x}(0)$ to the final state after one period, $\mathbf{x}(T)$.
By definition, Equation \ref{eq:gov} can be written in terms 
of the principal fundamental solution matrix,
\begin{equation}
 \frac{\mathrm{d}\boldsymbol{\Phi}_p}{\mathrm{dt}}=\mathbf{A}(t)\boldsymbol{\Phi}_p(t),
 \label{eq:matrix_equation_pfsm}
\end{equation}
and so $\boldsymbol{\Phi}_p(T)$ can be obtained directly by integrating 
Equation \ref{eq:matrix_equation_pfsm} forward in time one period.
The direct application of Floquet theory to a state vector describing 
a fluid flow  has been used by \citet{Noack94}, \citet{Robichaux99}, 
and \citet{Barkley96} 
to study instabilities in the 
periodic von K\'{a}rm\'{a}n vortex streets that develop in the wakes of cylinders.

\section{Floquet analysis of Mathieu's equation}\label{appB}
\begin{figure}
 \centering
 \includegraphics[width=0.7\textwidth,angle=0]{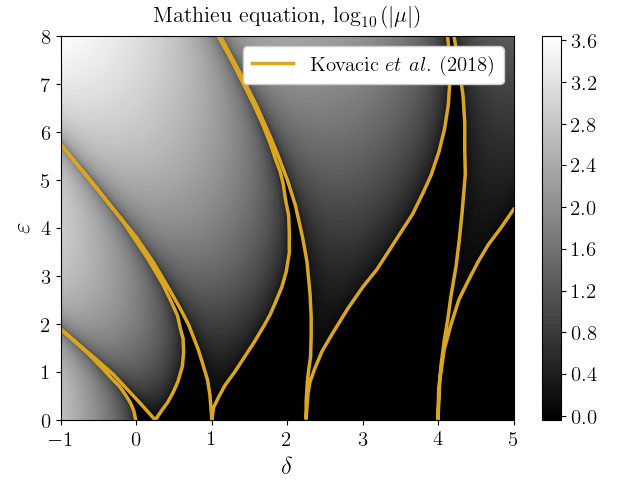}
 \caption{The neutral stability curve for Mathieu's equation.}
 \normalsize{\noindent Verification of the temporal discretization scheme and 
 the eigenvalue calculation. 
 The computed Floquet multipliers are shown by 
 the gray shading, and the $\Real[\mu]=1$ contour lies under the yellow line of 
 \citet{Kovacic18} to graphical accuracy. }
 \label{fig:mathieu}
\end{figure}

The Mathieu equation is an ordinary differential equation of the 
form
\begin{equation}
    \partial_{tt}y+f(t)y=0,
\end{equation}
where
\begin{equation}
    f(t)=\delta+\varepsilon\cos(t).
\end{equation}
The neutral stability curve of Mathieu's equation was computed to 
verify the temporal discretization and is shown in Figure \ref{fig:mathieu}.

\section{Finite difference grid convergence}\label{appC}
\begin{figure}
 \center
 \includegraphics[width=0.7\textwidth,angle=0]{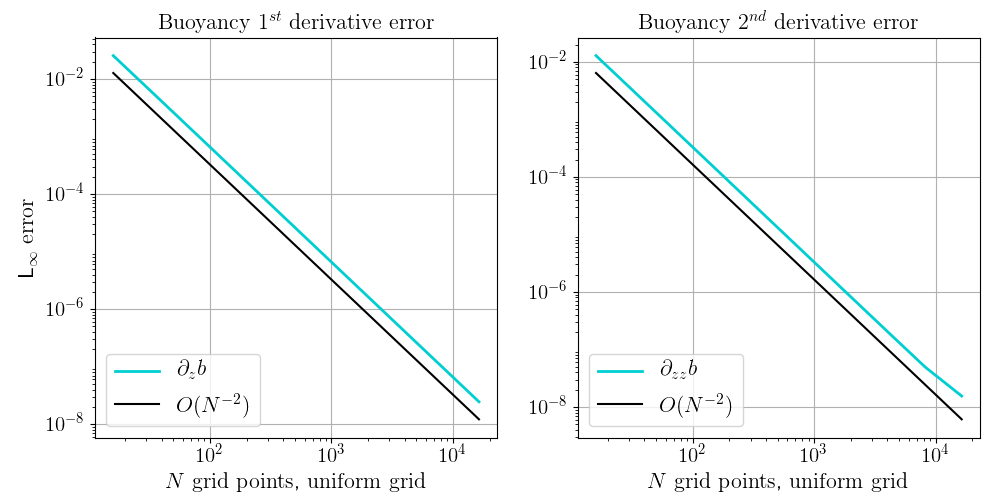}
 \caption{Grid convergence of buoyancy derivatives.}
 \label{b_grid_convergence}
\end{figure}

\begin{figure}
 \includegraphics[width=1\textwidth,angle=0]{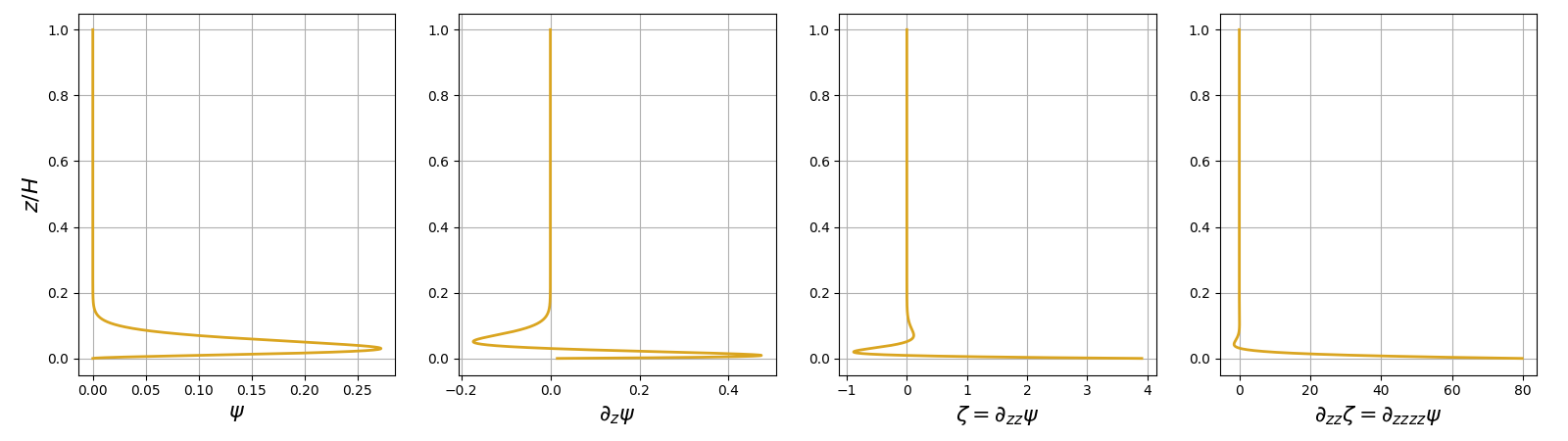}
 \caption{Test function for $\psi$.}
 \label{psi_test}
\end{figure}

\begin{figure}
 \includegraphics[width=1\textwidth,angle=0]{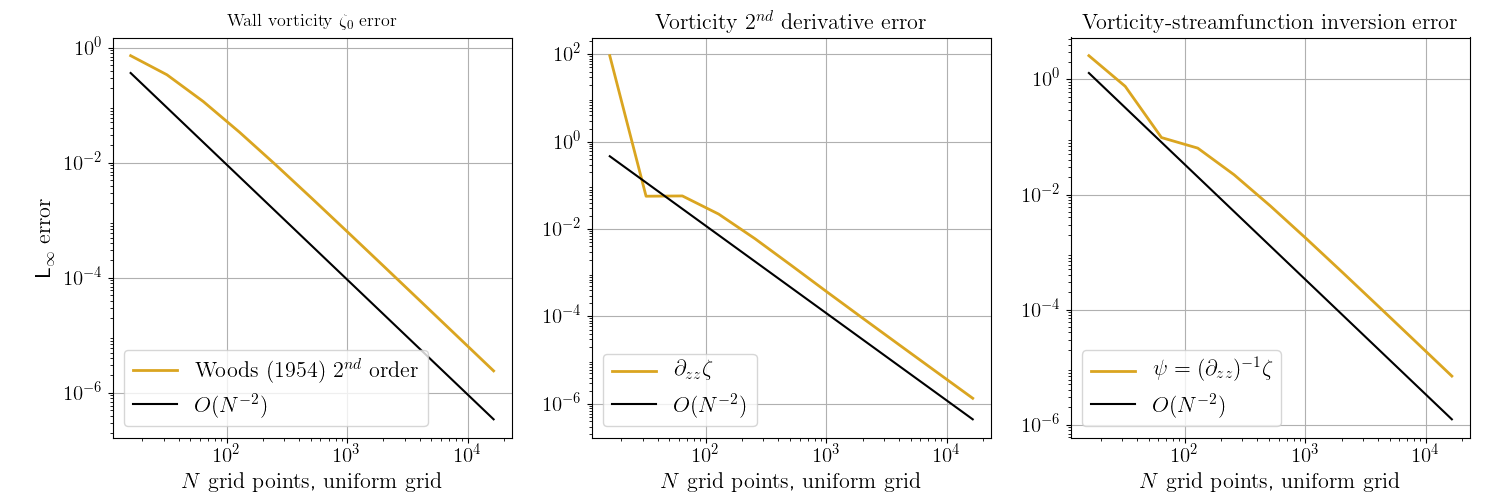}
 \caption{Grid convergence of finite differences for the vorticity.}
 \label{psi_grid_convergence}
\end{figure}

Second order accurate finite difference stencils were 
used to form discrete matrices 
for the calculation of the first and second derivatives 
of the buoyancy disturbances 
(Equation \ref{eq:buoyancy_zeta1}), 
for the calculation of the second derivatives
of the vorticity disturbances (Equation \ref{eq:streamwise_zeta1}),
and for calculating the streamfunctions by inverting the 
vorticity (Equation \ref{eq:inversion1}).
Figure \ref{b_grid_convergence} shows the grid convergence of 
the buoyancy derivative stencils when applied to the 
test function $b=\cos(2\pi{}z/H)$. 
The test function for checking the inversions and vorticity derivatives,
\begin{equation}
 \psi(z)=\big((z+1)^3-z^2-3z-1\big)\mathrm{e}^{-mHz}
\end{equation}
was chosen because it satisfies the same boundary conditions as 
were required for the Floquet analysis. The test $\psi(z)$ 
is shown in Figure \ref{psi_test}. The 
grid convergence of the Woods (1954) vorticity boundary condition 
(which imposes no-slip and impermeable boundary conditions on the diffusion of
vorticity), 
the second derivative of vorticity, and the inversion of vorticity 
to obtain the streamfunction are shown in Figure \ref{psi_grid_convergence}.

\section{Stokes' second problem governing equations}\label{appD}
The governing equation for the spanwise vorticity disturbances are
\begin{equation}
 \partial_t\zeta_2
 =\underbrace{\frac{\big(
 \partial_{zz}-k^2\big)}{2}{{\zeta}_2}}_{\text{diffusion}}
 -\underbrace{\frac{Uik\Rey}{2}{{\zeta}_2}}_{\substack{{\text{mean}}\\ {\text{advection}}}}
 +\underbrace{\frac{(\partial_{zz}U)ik\Rey}{{2}}{\psi}}_{\substack{{\text{vorticity line tilting}}\\ {\text{by mean flow}}}}
 +\underbrace{\textit{C}^2\big(\partial_{z}
 -ik\cot\theta\big){b}}_{\substack{{\text{baroclinic}}\\ {\text{production of vorticity}}}},
 \label{eq:spanwise_zeta2}
\end{equation}
and the governing equation for the buoyancy disturbances in 
the $x-z$ plane are
\begin{equation}
 \partial_t{b}=
 \underbrace{\frac{\big(
 \partial_{zz}-k^2\big)}{2\Pran}b}_{\text{diffusion}}
 -\underbrace{\frac{Uik\Rey}{2}b}_{\substack{{\text{mean}}\\ {\text{advection}}}}
 +\underbrace{\frac{(\partial_zB)ik
 \Rey}{{2}}{\psi}}_{\substack{{\text{advection of}}\\ {\text{mean buoyancy}}}}.
 \label{eq:buoyancy_zeta2}
\end{equation}
Equation \ref{eq:buoyancy_zeta2} and 
the baroclinic vorticity term in Equation \ref{eq:spanwise_zeta2}
are eliminated for tokes' second problem, where $\textit{C}=0$ and $b(z,t)=0$. 
The remaining spanwise vorticity disturbance equation (\citet{Blennerhassett06})
was used calculate the linear stability of Stokes' second problem shown 
in Figure \ref{fig:stokes_curve}.

\bibliographystyle{jfm}
\bibliography{jfm-instructions}

\end{document}